# Stretching the Printability Metric in Direct-ink Writing with Highly Extensible Yield-Stress Fluids

Chaimongkol Saengow[a,b,*], Samya Sen[b,†], Joaquin Yus[c], Eliza E. Lovrich[e],
Amanda G. Hoika[e], Kelly M. Chang[a,f], Arielle A. Pfeil[b], Nellie Haug[b,c],
Amy J. Wagoner Johnson[a,b,c,d,e,g,*], Randy H. Ewoldt[a,b,*]

[a] Beckman Institute for Advanced Science and Technology, University of Illinois Urbana-Champaign, Urbana, IL;

[b] Department of Mechanical Science and Engineering, The Grainger College of Engineering, University of Illinois Urbana-Champaign, Urbana, IL;

[c] Carl R. Woese Institute for Genomic Biology, University of Illinois Urbana-Champaign, Urbana, IL;

[d] Carle Illinois College of Medicine, University of Illinois Urbana-Champaign, Urbana, IL;

[e] Department of Bioengineering, The Grainger College of Engineering, University of Illinois Urbana-Champaign, Urbana, IL;

[f] Department of Materials Science and Engineering, The Grainger College of Engineering, University of Illinois Urbana-Champaign, Urbana, IL

[g] Chan Zuckerberg Biohub Chicago, LLC, Chicago, IL

## ABSTRACT

Direct-ink writing leverages the rheological complexity of yield-stress fluids to construct complex geometries, particularly those with large gaps across internal structures. However, extensional rheological properties have rarely been considered in work that studies rheology-printability correlations. Here, we test our hypothesis that extensional properties correlate with *drawability*, a key indicator of printability that signifies speed robustness, printing resolution, and gap-spanning performance. We formulated cementitious suspensions using hydroxyapatite (HAp) particles, independently tuning them for yield stress and extensibility – two crucial rheological properties – and test-printed. To enhance extensibility, we incorporated hydroxypropyl methylcellulose as a polymeric modifier, but this enhancement may decrease as yield stress increases, presenting a challenge in materials design. We modulated particle interactions to achieve a wide range of yield stress and extensibility, allowing for rigorous testing of our hypothesis. This approach created inks with high extensibility and high yield stress – generally considered mutually exclusive properties. We evaluated correlations between drawability and key rheological properties, finding the strongest positive correlation with extensional failure strains (strain-to-break) rather than yield stress. We establish a bijective property-manufacturing relationship (one-on-one mapping of shear yield stress to buildability and extensional strain-to-break to drawability) by combining our findings on drawability with previous studies on buildability. This relationship provides a comprehensive framework for designing high-performance inks that can be self-supporting, capable of high-speed printing, and allow gap-spanning features.

*corresponding authors (saengow@illinois.edu; ajwj@illinois.edu; ewoldt@illinois.edu)
†Current address: Department of Materials Science and Engineering, Stanford University, Stanford, CA.

# I  Introduction

Direct-ink writing (DIW), also known as direct-write assembly,[1,2] micro-robotic deposition,[3] and robocasting,[4] is an extrusion-based additive manufacturing technique that utilizes the rheological complexity[5–7] of extruded yield-stress fluid inks to construct complex geometries. DIW can excel in complex builds involving large-gap spanning ($L/D \gg 1$, where $L$ is gap distance and $D$ is filament diameter), enabling applications across various scales and disciplines. Examples include scaffolds for bone regeneration,[2,8] arched silver electrodes for LED pixels,[9] and free-form printing[2,8] (Figure 1A). These applications demonstrate the versatility of DIW in producing structures with significant gap-spanning capabilities.

To optimize materials for DIW, it is crucial to understand the concept of printability, which can be broken down into several key components based on the targeted applications.[10–13] Fu *et al.* offered a comprehensive analysis of this equivocal metric[12], which we define in this work as a synthesis of *buildability* and *drawability*. Figure 1B illustrates the two failure modes associated with these characteristics. While the existing definition of printability only addresses buildability, we propose an expanded concept that integrates both. Buildability represents the maximum gravitational stress the building materials can withstand before collapsing. While shear yield stress ($\sigma_y$) plays a critical role in determining buildability by allowing the material to resist more weight, other factors, such as linear elastic modulus ($G'$), also contribute. Milazzo *et al.* further demonstrated that, for similar $\sigma_y$, buildability favors higher $G'$ and lower yielding shear strain ($\gamma_y$).[11]

While buildability is essential for structure integrity, drawability is critical in high-speed printing, achieving complex geometries, and gap-spanning capabilities. Drawability, derived from polymer fiber spinning,[14] is crucial in direct-ink writing for complex gap-spanning builds. It enables the formation of straight, cylindrical filaments when stretched, ensuring high shape fidelity. Yuk and Zhao studied viscoelastic inks and harnessed drawability for higher printing resolution, though without detailed rheological analysis.[15] Nelson *et al.* were the first to consider imparting extensibility (using extensional strain-to-break as a metric) to yield-stress fluids, laying the groundwork for further research.[16] Building on this, Rauzan *et al.* showed that inks with higher extensibility increases the ability of filaments to span relatively large gaps.[17]

Understanding drawability requires comprehension of the different regimes that occur during printing. Dispensers can control two relevant speeds: nozzle speed ($V_{nozzle}$) and average ink speed ($V_{ink}$), with their ratio defining the draw ratio (DR $\equiv V_{nozzle}/V_{ink}$), a metric used to quantify drawability in this work. Figure 1C illustrates four DR-governed regimes: over-extrusion (DR < 1), speed matching (DR = 1), drawdown (DR > 1), and filament breakup (DR > $DR_b$). Over-extrusion causes a larger extrudate diameter if building on a substrate and sag if spanning across a gap. The speed matching regime produces extrudate matching nozzle diameter. The drawdown regime enables higher resolution and large gap-spanning. The at-break draw ratio ($DR_b$) represents the drawability limit of the material.

Building on this understanding of printability and its components, we propose a hypothesis that links material properties to printing performance. Two fundamental rheological properties in DIW are yield stress, which is crucial for buildability, and extensibility, whose effects are less understood. Figure 1D illustrates these four regimes within the printing-parameter space ($\Delta P - V_{nozzle}$), where the pressure drop, $\Delta P$, controls $V_{ink}$. At a fixed $\Delta P$ (or $V_{ink}$), increasing $V_{nozzle}$ raises the draw ratio. Previous studies hint at a positive correlation between extensibility and gap-spanning performance.[17] This clue led to our hypothesis that material extensibility can broaden the drawdown regime, as shown in Figure 1D.

While many DIW studies have measured rheological properties, few have thoroughly analyzed their correlation with printability.[18–20] This paper addresses this gap by exploring the relationships between printability and two critical properties—shear yield stress ($\sigma_y$) and extensional engineering strain-to-break ($\epsilon_{STB}$)—using dense hard-particle suspensions. By investigating these connections, we aim to enhance the understanding and optimization of DIW materials.

## II    Microstructural Tuning Mechanisms for Yield Stress and Extensibility

This study presents the first exploration of extensible yield-stress fluids[16] using hard-particle suspensions, advancing beyond the established knowledge of soft-particle suspensions.[17,21] Hard-particle suspensions present unique challenges in tuning yield stress due to their high sensitivity to volume fraction. Unlike soft-particle suspensions, where yield stress primarily arises from Hertzian-like interactions between compressed bodies above random close packing condition,[22–26] hard-particle systems are more sensitive to shape irregularities and friction,[27] and surface chemistry,[28–30] affecting interparticle attractions or repulsions that generate yield stress. Despite these tuning challenges, hard particles remain crucial for achieving a broader yield-stress spectrum, albeit with potentially reduced extensibility.

We focus on hydroxyapatite (HAp) as our primary material, given its ability to form cementitious pastes with yield stress through hydration[31] and its excellent candidacy for implantable bone scaffolds. HAp can gradually remodel into new bone post-implantation, eliminating the need for surgical removal.[31] Research indicates that well-fabricated scaffolds with minimal sag-induced errors promote better bone growth through capillary flow,[8] with an optimal lattice size of approximately 500 $\mu$m.[2] These fabrication requirements motivate our development of complex fluids that simultaneously exhibit high extensibility and high shear yield stress, aiming to create stronger constructs for various applications.[8,9,32,33]

### a    Yield-Stress Fluids

Yield stress is one of the crucial properties in direct-ink writing with complex fluids, originating from having a load-bearing network spanning the entire system. A yield-stress material resists deformation like a solid until it exceeds a critical yield stress, after which it flows like a fluid. Yield stress in fluids may arise from one of the two distinct microstructural mechanisms: (i) repulsion-dominated and (ii) attraction-dominated, although sometimes, both mechanisms may contribute simultaneously.[22–24] Experimental observations show that these two extremes have distinct rheological signatures: repulsion-dominated fluids offer wide-ranging yield stress but rupture easily in extension, while attraction-dominated fluids have lower yield stress but higher ductility.[23]

Our HAp-based inks consist of three main components in an aqueous environment: hydroxyapatite particles (HAp), polymethyl methacrylate microspheres (PMMA), and hydroxypropyl methylcellulose polymer (HPMC, a modified cellulose[34]), as shown in Figure 2A (also Figure S7– S12).  If HAp particles are in close proximity, they can aggregate through hydration[31] and mutate ink rheology, quickly altering the flow properties.[35] To manage this, throughout this work, we incorporate an inert phase of PMMA at a 1:1 volume ratio with HAp, which inhibits the hydration reaction. However, careful control of PMMA loading remains essential, as excessive amounts can compromise the final construct performance.[36] Figure 2B confirms that PMMA particles are larger than HAp, explaining how PMMA effectively inhibits the HAp hydration reaction. Whereas our synthesized HAp can readily form networks, resulting in yield stress, the inert PMMA does not, even approaching random close packing conditions at 63 vol% (see the tilt tests as protorheology evidence[37,38] in Figure 2C). The HAp and PMMA particle sizes and distributions reported in Figure 2B show that both particles are large enough and in the dense suspension regime that renders them athermal, where particle diffusion due to thermal effect is negligible (Péclet number: Pe $\equiv 6\pi r^3 \sigma_c / kT \gg 1$). However, both particles are still small enough that colloidal physics – including steric effects and electrical double-layer forces[18] – plays a significant role, as evident in Figure 2C.

Incorporating polyelectrolytes to the inks is another tuning knob to modulate yield stress.[28,30] This study employs two polyelectrolytes, polyacrylic acid (PAA) and polyethylenimine (PEI), which can adsorb onto HAp *via* hydrogen bonds.[39,40] Deprotonated PAA can adsorb onto HAp surfaces and increase negative charge, leading to higher interparticle repulsions. Consequently, PAA addition can enhance dispersion and reduce yield stress by weakening the microstructural network. Figure 2C shows that PAA can lower the zeta potential to -50 mV. Subsequently, PEI addition as a second layer can raise the zeta potential to +40 mV, partially restoring the network and yield stress. Our results indicate that PMMA is inert to surface treatment as the zeta potential is relatively constant with respect to both PAA and PEI contents (Figure 2C). Therefore, our modifications only alter HAp surfaces.

The mechanism behind this yield-stress modulation *via* surface treatment can be electrostatically explained by the DLVO theory, as illustrated in Figure 2D. The lowest repulsion potential energy allows neat HAp particles to achieve the closest interparticle distance. PAA addition increases this repulsion, pushing particles apart. Subsequent PEI addition reduces this energy, bringing particles closer. The attractive van der Waals force profile remains unchanged throughout these modifications. This interplay governs interparticle spacing and thus modulates yield stress. This proximity modulation allows us to create inks with a broad spectrum of rheological properties to test our hypothesis, as stated in Figure 1D.

To articulate our ink compositions and treatments clearly, we define the volume fraction of the $i$th species as $\phi_i = V_i / (V_{HAp} + V_{PMMA} + V_w)$. This definition applies to our suspensions of HAp and PMMA in deionized water, with HAp and water forming the paste and PMMA serving as inert inclusions, allowing us to consider the overall particle fraction as well as sub-component volume fractions, which is important for identifying different packing conditions. We use the following format to designate our samples: H$xx$-S$yy$-C, where $xx$ represents the milligrams of HPMC per milliliter of ink (overall volume), $yy$ denotes the total volume fraction of solid particles (HAp and PMMA at 1:1 vol%) in the ink, and the suffix -C indicates HAp surface treatment with 1 wt% PAA and PEI. For instance, H10-S55-C means an ink with 10 mg HPMC/mL, solid particle volume fraction $\phi_p \equiv (V_{HAp} + V_{PMMA}) / (V_{HAp} + V_{PMMA} + V_w) = 55\%$ made with PAA-PEI surface-treated HAp.

### b   Extensible Yield-Stress Fluids

Yield-stress fluids can be highly extensible.[16,21] Such extensibility contradicts earlier predictions based on tensorial viscoplastic models like Bingham and Herschel-Bulkley. These traditional models, which describe shear properties well, coupled with the Considère criterion, imply that filaments in extension should be unstable to necking at any extensional strain rate.[16] However, real-world observations challenge these predictions, as extensibility is readily observed and engineered in various yield-stress fluids,[16] such as high internal phase emulsions[17] and dense microgel suspensions.[21]

Nevertheless, imparting extensibility to dense hard-particle suspensions remains challenging. This difficulty arises from the high sensitivity to surface interactions of microstructural networks in hard-particle suspensions. In weaker networks, yield stress and extensibility correspond closely, as the yield stress mitigates the destabilizing effects of Laplace pressure from surface tension.[16,21,41] However, this relationship may break down in a stronger network,[16] due to several factors such as sample heterogeneity and rate-dependency from viscoelasticity (Deborah number $\gtrsim 1$).[42] These complexities motivate our research to develop highly extensible, high yield-stress materials by carefully modulating microstructural mechanisms, aiming to overcome the current limitations in dense hard-particle suspensions.

It is crucial to examine the molecular mechanisms that enhance extensibility to understand how we can create extensible yield-stress fluids. One way is by giving rise to the extensional viscosity to resist breakage caused by an instability such as necking[43,44] or surface tension (capillarity) driven breakup.[19] Filament stretching extensional rheometry (FiSER) can measure this extensibility by applying a controlled stretching speed ($V_{\text{stretch}}$) and examining the breakup behaviors and the thinning speed at the narrowest point (see Figure 3B and Figure S2). The extensional strain-to-break ($\epsilon_{\text{STB}}$) depends on $V_{\text{stretch}}$ and $V_{\text{ink}}$, with the latter is influenced by fluid properties (surface tension, $\Gamma$, and density, $\rho$), geometry (initial sample height, $h_0$, and diameter, $D_0$), and rheological properties (relaxation time, $\tau$, characteristic viscosity, $\eta_C$, and yield stress, $\sigma_y$).

Although $\epsilon_{\text{STB}}$ is an extrinsic metric due to its geometrical dependence, it remains the most suitable metric to measure sample extensibility across a vast regime, from complex fluids[16,17,21,23,42] to solid[45] materials. The strain-to-break is particularly practical for materials whose necking is less pronounced or challenging to locate, such as our HAp-based inks. It also offers a visual connection to actual manufacturing processes, rationalizing our choice of $\epsilon_{\text{STB}}$ as a quantification metric for extensibility.

The thinning speed is related to fluid properties, including viscosity, yield stress, density, and surface tension (Figure 3B),[16,19,21,23] leading to at least three key dimensionless numbers: the Ohnesorge number (Oh $\equiv \eta_C/\sqrt{\rho \Gamma R}$), the capillary number (Ca $\equiv \eta_C \dot{\epsilon}_c R/\Gamma$), and the Bingham number (B $\equiv \sigma_y/\eta_C \dot{\epsilon}_c$), where $\eta_C, \sigma_y, \rho, \Gamma, R, \dot{\epsilon}_c$ are characteristic viscosity, yield stress, density, surface tension, instantaneous filament radius, and characteristic extensional rate. Understanding these numbers is essential for creating highly extensible yield-stress fluids.

Figure 3A compares the stretching behavior across a wide range of formulations, providing initial observations on extensibility. As a comparative control, DI water forms a stable capillary bridge at a short length, and extending beyond this stable length causes the filament to break quickly due to low viscosity. Raising the characteristic viscosity with a polymer (HPMC solution) can extend the breaking strain further by slowing down the necking processes. HAp-based materials show increasing capillary-gap stability as particle fraction rises, likely due to the enhanced yield stress, balancing capillary pressure, which is captured by the plasto-capillary number (Ca$_P$ = Ca × B $\equiv \sigma_y R(t)/\Gamma$). However, complex relationships emerge between particle and HPMC loadings and extensibility. Extensibility drops once a threshold is reached (H0-S50). Surprisingly, adding polymers, typically used to enhance extensibility, produces a detrimental effect in this case (H5-S50). These observations challenge the presumption, suggested by Ca$_P$, that higher yield stress should lead to improved extensibility. Despite these apparent contradictions, we find that with careful formulation, it is possible to achieve both high yield stress and high extensibility, as demonstrated by H10-S55-t. This sample exhibits significantly higher elongation before breaking, suggesting that the relationship between composition and extensibility is more nuanced than initially apparent.

To understand these complex behaviors, it is crucial to examine how extensibility increases in a viscoelastic medium, precisely the HPMC solutions. Figure 3C illustrates the zero-shear viscosity of HPMC in DI water across the concentration range used in our HAp-based inks. The Ohnesorge number spans from approximately 0.8 to 800, indicating a transition from inertio-capillary to visco-capillary breakup regimes, which happens when Oh ~ 1.[5,46,47] This transition is expected to be even more pronounced with particle addition and the accompanying increases in viscosity, allowing us to exclude inertial effects from our analysis. Our focus then shifts to the remaining two key dimensionless numbers: the capillary number (Ca) and the Bingham number (B), whose product yields the plasto-capillary number ($Ca_P$). These parameters become crucial as we transition from inertio-capillary to visco-capillary breakup regimes. Ca predominantly governs the behavior in regimes where yield stress is small, typically at low particle volume fractions ($\phi_p <$ ~40% for our HAp-based materials). As $\phi_p$ increases beyond this threshold, B and $Ca_P$ become more pertinent.

When the particle fraction is intrinsically small, we consider only the continuum viscoelastic medium of HPMC in DI water. In this regime, we observe a monotonic increasing trend in extensional strain-to-break across the range of HPMC concentrations used to formulate our HAp-based inks. We observe the same trend for zero-shear viscosity. Co-plotting zero-shear viscosity and extensional strain-to-break reveals a strong linear and monotonic relationship in log-log space (Figure 3C). This relationship aligns with the concept of capillary number, suggesting that higher viscosity enhances extensibility before the pinch-off event occurs, even when the unstable capillary length may have been exceeded. Any non-monotonic relationship between shear and extensional properties that emerges upon the addition of hard particles can thus be attributed to the particle addition. These findings set the stage for a more detailed mechanistic discussion in the analysis of our HAp-based inks (Section IIIa), where we will explore the interplay between yield stress and extensibility more comprehensively.

## III     Results: Microstructure-Rheology-Printability Relationships

In DIW, the ability to stretch filaments without rupture is crucial for improved shape fidelity.[15] The plasto-capillary number can signify this ability, quantifying the balance of yield stress against Laplace pressure in thinning filaments (Figure 4A). This balance suggests that a higher yield stress could result in a thinner filament, resulting in higher strain-to-break. However, earlier observations suggested that increased yield stress does not necessarily enhance extensibility,[16,19] especially in hard-particle suspensions. Careful materials design can help to achieve both high extensibility and high yield stress for improved printability.

In Section IIIa, we examine the intricate relationship between yield stress ($\sigma_y$) and engineering extensional strain-to-break ($\epsilon_{STB}$). We explore how HPMC loading, particle volume fraction, and electrostatic surface modulation affect these two properties. Our investigation begins by examining how particle volume fraction and HPMC loading influence $\sigma_y$ and $\epsilon_{STB}$. An Ashby-style diagram, co-plotting these two properties, highlights a unique region exhibiting high extensibility and yield stress (see also Figure S14 and S15).

Building on this analysis, we design our HAp-based inks targeting this unconventional region by modulating HAp interactions using HPMC, PAA, and PEI, allowing for higher particle and HPMC loading. In Section IIIb, we evaluate the drawability of these inks by printing across a large gap and quantifying it using draw ratios. Finally, we investigate how $\sigma_y$ and $\epsilon_{STB}$ correlate with two critical draw ratios: the highest before break ($DR_\ell$) and the lowest observation after breakup ($DR_b$). This comprehensive approach allows us to establish robust relationships between material properties and printability in gap-spanning scenarios.

### a  Microstructure-Rheology

We start with the fundamental relationship between particle volume fraction and yield stress, focusing only on HAp and PMMA without either HPMC or PAA-PEI surface treatment. Figure 4B shows that increasing particle volume fraction ($\phi_p$) enhances yield stress ($\sigma_y$), suggesting a strengthened network.[2,11,48] We observed a significant increase in yield stress, starting even at the much lower volume fraction ($\phi_p$ ~40 vol%) than the random close packing, reflecting HAp hydration reactivity, in contrast to the inert nature of PMMA (as shown earlier in Figure 2C). At a 1:1 volume ratio of HAp-to-PMMA, particle loading reaches a maximum of approximately 53 vol%. Beyond this threshold, significant heterogeneity emerges, posing challenges for further increasing particle content. This behavior aligns well with other studies on dense hard-particle suspensions, where particle interactions and network formation primarily govern yield stress development.[2,11,17,21,49]

Building on this foundation, we explore the role of polymer addition, specifically HPMC, in enhancing yield stress. Polymer addition can enhance yield stress through increased steric stabilization, reported in both soft glassy[21] and hard-particle[50] suspensions. We observe the similar trend in our HAp-based materials as shown in Figure 4B. HPMC acts as a primary network modulator by providing steric stabilization between particles. Its ability to bind to HAp surfaces significantly enhances the steric energy, as shown in Figure 4B at lower HPMC loading. The effect of HPMC is more synergetic as $\phi_p$ increases, which may be attributed to more binding sites for HPMC.

Further modulation of the network strength can be achieved through PAA-PEI surface treatment by altering electrostatic interactions. PAA initially increases particle separation through electrostatic repulsion, thus weakening the network and resulting in a lower yield stress. At this point, particles are properly fluid-mediated, minimizing the stress concentration from depletion effects. The subsequent PEI addition reduces the repulsive force, allowing particles to get closer and thus partially rejuvenating the network strength. This interplay between HPMC-mediated effects and polyelectrolyte-controlled electrostatic forces enables us to modulate the microstructural network and resulting yield stress while maintaining good particle dispersion. As shown in Figure 4B, this wide range of yield stress from the same set of materials is remarkable and, to our knowledge, potentially unprecedented in its extent within the current literature.

Figure 4C reveals a non-monotonic trend in extensibility for hard particles, quantified using extensional strain-to-break, $\epsilon_{STB}$, from filament stretching extensional rheometry (FiSER) reflecting the interplay of two competing mechanisms: polymer bridging and particle crowding. Figure 2A hints at the crucial role of HPMC in enhancing extensibility by creating polymer bridges between particles through hydrogen bonding, especially at lower volume fraction. This bridging effect becomes more pronounced with increasing HPMC concentration, as evidenced by the higher strain-to-break observed in formulations with higher HPMC content (compare H0-S40 and H5-S40 in Figure 3A). As particle loading increases, it initially complements the HPMC in raising extensibility (compare DI-water and H0-S40 in Figure 3A). However, this synergistic enhancement reaches a limit at higher volume fractions, and extensibility starts to drop with particles and HPMC (compare H5-S40 and H5-S50 in Figure 3A).

To overcome this limitation, we employed electrostatic surface treatment using PAA and PEI. This strategy improves particle dispersion. We speculate that this PAA-PEI treatment allows for more uniform HPMC distribution throughout the system. This improved dispersion enables both higher polymer and particle loadings while maintaining effective polymer bridging. The subsequent PEI addition fine-tunes the repulsion potential energy, allowing particles to get closer while maintaining good dispersion. With careful electrostatic modulation, we show that H10-S55-C can exhibit higher extensibility, showing significant elongation before breaking. The result is a well-connected microstructural network since the interparticle spacing is small enough for the hydration reaction yet sufficiently separated for HPMC bridging. This balanced structure yields materials that exhibit high extensibility and high yield stress–the two rheological properties that are typically mutually exclusive, yet that are unified in this work.

We further provide protorheology as visual confirmation of rheological properties and their consequences for flow.[37,38] Extensibility is demonstrated through filament stretching until break at $t = t_b$, with higher $t_b$ indicating higher extensibility (Figure S17–S54). Shear yield stress is shown through superimposed images of the bottom of the sample after filament breakup, taken at $t = t_b$ and $t = 2t_b$. Samples with higher yield stress better retain their final shapes (Figure S55). These visual observations directly relate to buildability (directly related to yield stress) and drawability (directly related to strain to break).

The Ashby plot in Figure 4D, depicting yield stress against strain-to-break, reveals a microstructure-agnostic relationship that synthesizes our observations from Figures 4B and 4C. At low yield stress, we initially observe a close correspondence between these properties, reflecting the simultaneous increase in yield stress and the extensibility, attributed to either increasing viscous stress to slow down the pinch-off (increasing capillary number) or a more stable liquid bridge, as suggested by the plasto-capillary number. However, a complex relationship emerges between yield stress and extensibility. We learned from Figure 4B that yield stress continues to increase with particle loading, and we learned from Figure 4C that extensibility drops as a threshold is reached, maybe due to particle crowding or extensional viscous stress struggles to overcome yield stress and cannot drive the flow (B ≫ 1). This flipped trend contradicts the simple force balance suggested by the plasto-capillary number, which suggests that higher yield stress should enable a smaller filament before break (thus higher extensibility). Overall, for our HAp-based materials, yield stress inversely related to strain-to-break, aligning with trends observed in other yield stress fluids.[16,23] This trend underscores the trade-off between good shape retention and extensibility in particulate systems.

The most important observation from Figure 4D is the effect of the PAA-PEI surface treatment. In contrast to untreated samples, yield stress of surface-modified materials is directly related to strain-to-break across a wide range of values. This remarkable trend validates the efficacy of our tuning protocol in creating fluids with simultaneously high yield stress and extensibility. This directional relationship stems from improved particle dispersion achieved through surface treatment. The PAA-PEI treatment enables higher polymer and particle loadings while minimizing the particle crowding effect. The resulting well-connected network, with well-spaced particles, exhibits high shape retention (high yield stress) while accommodating large deformations before failure (high extensibility).

### b  Rheology-Printability

Having established the microstructure-rheology relationships in the previous section, this section explores how rheological properties link to printability, specifically drawability in gap-spanning scenarios. DIW of self-supporting structures with gap-spanning features, such as lattices, is challenging.[51] The filament must maintain tension while stretching; drawability can quantify this behavior. To evaluate drawability, we conducted a controlled gap-spanning test (details in Section SIII.d). By keeping the average ink speed ($V_{ink}$) constant and increasing nozzle speed ($V_{nozzle}$), we progressively raised the draw ratio (DR $\equiv V_{nozzle}/V_{ink}$, Figure 5A). Filament breakup occurs when DR exceeds the drawability limit. We can quantify drawability using two draw ratios: the last successful-point draw ratio ($DR_\ell$) and the one at-break ($DR_b$). While $DR_b$ indicates maximum drawability, $DR_\ell$ reveals how well the ink can span gaps before breaking, which is crucial for manufacturing complex lattice structures with high shape fidelity.

Figure 5B illustrates the relationship between draw ratios, particle volume fraction, and HPMC loading. Generally, draw ratios decrease with increasing particle volume fraction. However, PAA-PEI-treated samples show the opposite trend. Notably, H10S50-C exhibits a significantly higher draw ratio, mirroring extensional strain-to-break trends. This observation hints at a strong correlation between draw ratio and extensibility, which we explore further.

To connect filament stretching during DIW with FiSER, we analyze the extrudate kinematics in two interconnected scenarios, both occurring over the same spanning time $t_{span}$. In the first scenario, extrusion occurs without stretching, rendering it in the speed matching regime, where $V_{nozzle} - V_{ink}$, resulting in a filament of length $L_0$. This establishes a baseline for strain calculation, analogous to the initial state in FiSER. The second scenario represents typical spanning in the thinning regime, where the stretching speed is defined as $V_{stretch} = V_{nozzle} - V_{ink}$, enabling successful gap spanning across length $L$ (see also Figure S16). This second scenario mirrors the stretching process in FiSER. By framing both scenarios within the context of the same spanning time, we can use the spanning lengths from each scenario ($L_0$ and $L$) for strain calculation.

For a simplified case of uniform uniaxial extension under quasi-steady state conditions (where the Deborah number De $\equiv \tau/t_{span} \ll 1$, where $\tau$ is the relaxation time), similar to that of a viscous catenary,[52] we express the extensional strain associated with the two relevant speed ($V_{nozzle}$ and $V_{ink}$) and the gap length ($L$) as:

$$\epsilon_{viscous} = \frac{L}{L_0} = \frac{(V_{nozzle} - V_{ink})t_{span}}{V_{ink}t_{span}} = \frac{V_{nozzle}}{V_{ink}} - 1 = DR_{viscous} - 1 \qquad (1)$$

where $L$ and $L_0$ are from the thinning and speed matching scenarios, described previously. This relationship establishes a maximum theoretical limit line under the aforementioned assumptions. It predicts the upper bound for extensional strain (equivalent to the strain to break from FiSER) and the draw ratio (from DIW). Our experimental observations in Figure 5C align well with this limit.

This observation is crucial to our findings. Statistical analysis confirms strong correlations between $\epsilon_{STB}$ and both $DR_\ell$ and $DR_b$, with the Pearson coefficients ($\rho_P$) of 0.93 and 0.92, and the Spearman coefficients ($\rho_S$) of 0.88 and 0.85, indicating robust monotonic relationships (Figure 5C). In contrast, the analysis of Figure 5D reveals that yield stress alone is insufficient to quantify drawability. Yield stress presents significantly weaker correlations ($\rho_P = 0.27, 0.26$), making it a less reliable predictor of drawability. Consequently, we can exclude yield stress from the rheology-drawability relationships. Using linear regression, we report the relationships between the two draw ratios and $\epsilon_{STB}$:

$$DR_\ell = 0.489\epsilon_{STB} + 0.512; DR_b = 0.537\epsilon_{STB} + 0.534 \qquad (2)$$

These fitted relationships provide a predictive tool for drawability based on measured $\epsilon_{STB}$. Importantly, extensional properties are often overlooked in conventional analyses; without considering them in this work, we would not have identified such a strong correlation.

## IV    Conclusion

This study demonstrates how particulate ink printability can be enhanced in direct-ink writing (DIW) by incorporating extensibility into hard-particle suspensions. We expand the concept of printability to include drawability, addressing critical aspects of constructing complex structures such as speed robustness, high-resolution printing, and gap-spanning. Drawability is crucial as it enables the extrudate to maintain tension during high-speed printing, facilitating greater mismatches between average ink speed and nozzle movement speed. Using a gap-spanning experiment, we quantified drawability using an at-break draw ratio by increasing nozzle speed until filament breakup occurred. Using hydroxyapatite-based cementitious inks with varying yield stress and extensibility, we confirm a strong correlation between drawability and extensional strain-to-break with the Pearson coefficient, $\rho_P$, of 0.93 and 0.92.

Our tuning protocol modulated particle interactions, achieving a broad range of yield stress (18.2–1240 Pa) and extensibility (extensional strain-to-break of 68.5–719%), surpassing the typical ranges seen in a single material system (illustrated in the Ashby-style plot along with other materials in Figure S14). This well-tuned system also achieved a distinctive combination of properties with both high yield stress and high extensibility. Statistical analysis reveals extensional strain-to-break as a reliable predictor of drawability, demonstrating a strong linear and monotonic relationship while showing only a weaker link with shear yield stress. This finding underscores the critical role of extensional rheology – an often-overlooked property in conventional analyses – in identifying this strong correlation.

Our research has established a predictive relationship between draw ratios and extensional strain-to-break, while previous studies have already linked yield stress to buildability. These two correlations form a bijective property-manufacturing relationship (*i.e.*, one-on-one mapping of shear yield stress to buildability and extensional strain-to-break to drawability) that is applicable to any extrusion-based additive manufacturing, such as bioprinting[53] and fused filament deposition.[54] Importantly, by considering both shear yield stress and extensibility, we provide a more comprehensive metric for printability to design high-performance inks that can be self-supporting, capable of high-speed printing while allowing gap spanning features.

## V    Supporting Information

The data that support the findings of this study are available in the supplementary material. Complete dataset is available upon request from the authors.

## VI    Acknowledgment

CS acknowledges postdoctoral support from the Beckman Postdoctoral Fellows Program and the University of Illinois Urbana-Champaign; Prof. Simon A. Rogers for linking LAOStress timescale to filament breakup; Prof. Julian C. Cooper for chemistry discussion. The PMMA microspheres were kindly donated by Tetsuya Okano at Sekisui Plastics Co., Ltd. (Japan). The authors are grateful to Anton Paar for loaning MCR 702 rheometer to the Ewoldt Research Group for all shear rheological measurements. A. Wagoner Johnson is a CZ Biohub Investigator.

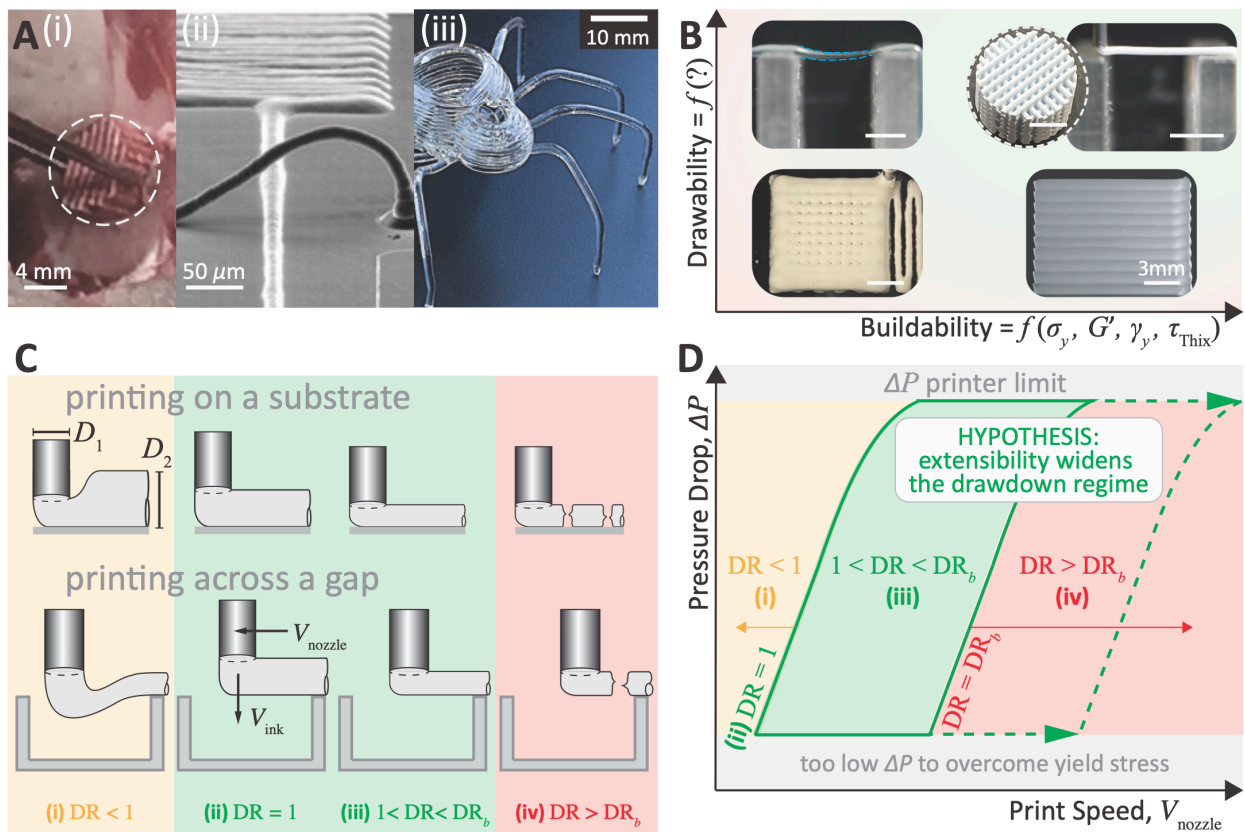

**Figure 1: Hypothesized Effect of Extensibility on Printability. A.** Complex builds requiring high drawability: (i) scaffolds for bone regeneration[8]; (ii) arched silver electrode[9]; (iii) free-form printing.[6] **B.** Two key aspects of printability: buildability and drawability. Buildability is influenced by yield stress, elastic modulus, and yield strain, while drawability, a novel concept introduced in this work, governs gap-spanning performance. Inks with high drawability and buildability enable successful layer stacking and large gap bridging. Scale bars: 3mm. Lower right example from M'Barki *et al.*[55] **C.** Draw ratio, $DR \equiv V_{nozzle}/V_{ink}$, determines four printing regimes both for on-substrate printing and printing that spans a gap: (i) over-extrusion ($DR < 1$), (ii) speed matching ($DR = 1$), (iii) drawdown ($1 < DR < DR_b$), and (iv) discontinuous, $DR > DR_b$. **D.** Printing parameter space of pressure drop ($\Delta P$) and nozzle speed ($V_{nozzle}$) with the four regimes, showing our main hypothesis that extensibility widens the drawdown regime, allowing for high drawability.

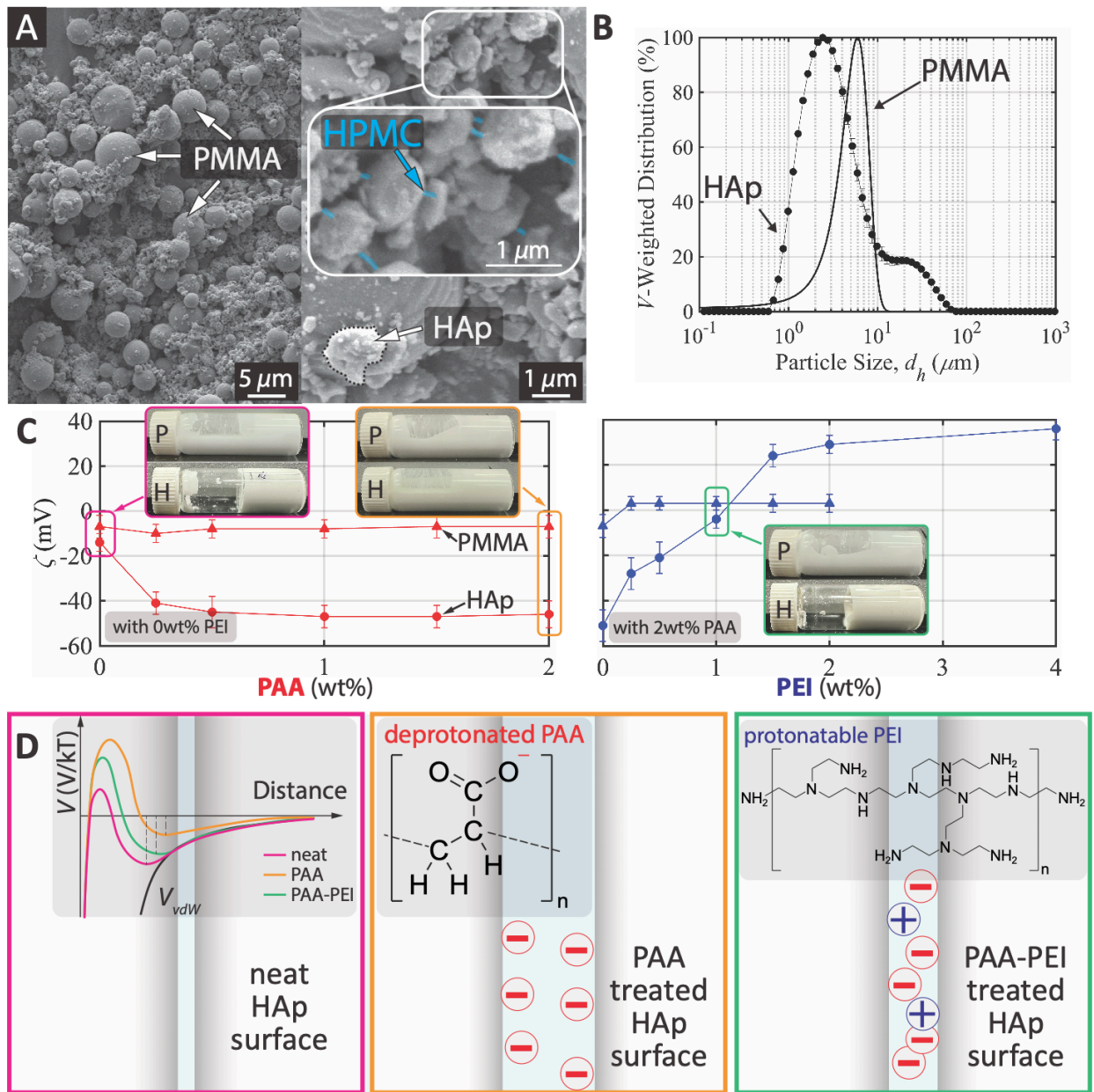

**Figure 2: Material design: microstructure-rheology tuning mechanisms of yield stress. A.** SEM micrograph of 1:1 vol% HAp:PMMA sample with 5wt% HPMC shows the distribution of irregularly shaped HAp, PMMA microspheres and HPMC fibrils (representatives in cyan). **B.** Volume-weighted particle size frequency distribution of PMMA (from Ref. [56]) and HAp. **C.** Zeta-potential of PAA-treated (red) and PAA-PEI-treated (blue) HAp and PMMA. Agglomeration occurs when $\zeta \rightarrow 0$ increases the yield stress. Sixty-second tilt test[38] using a vial with internal diameter of 15 mm shows signature of high $\sigma_y$ for neat and PAA-PEI-treated HAp (50 vol%), while PMMA at ~random close packing condition (63vol%) and PAA-treated HAp shows no sign of $\sigma_y$ and flows within two seconds after tilting. Error bars: one standard deviation. **D.** Interactions between two HAp particles for three surface treatments. The sketch of DLVO interaction energy profiles illustrates PAA and PEI dispersion effects on HAp.

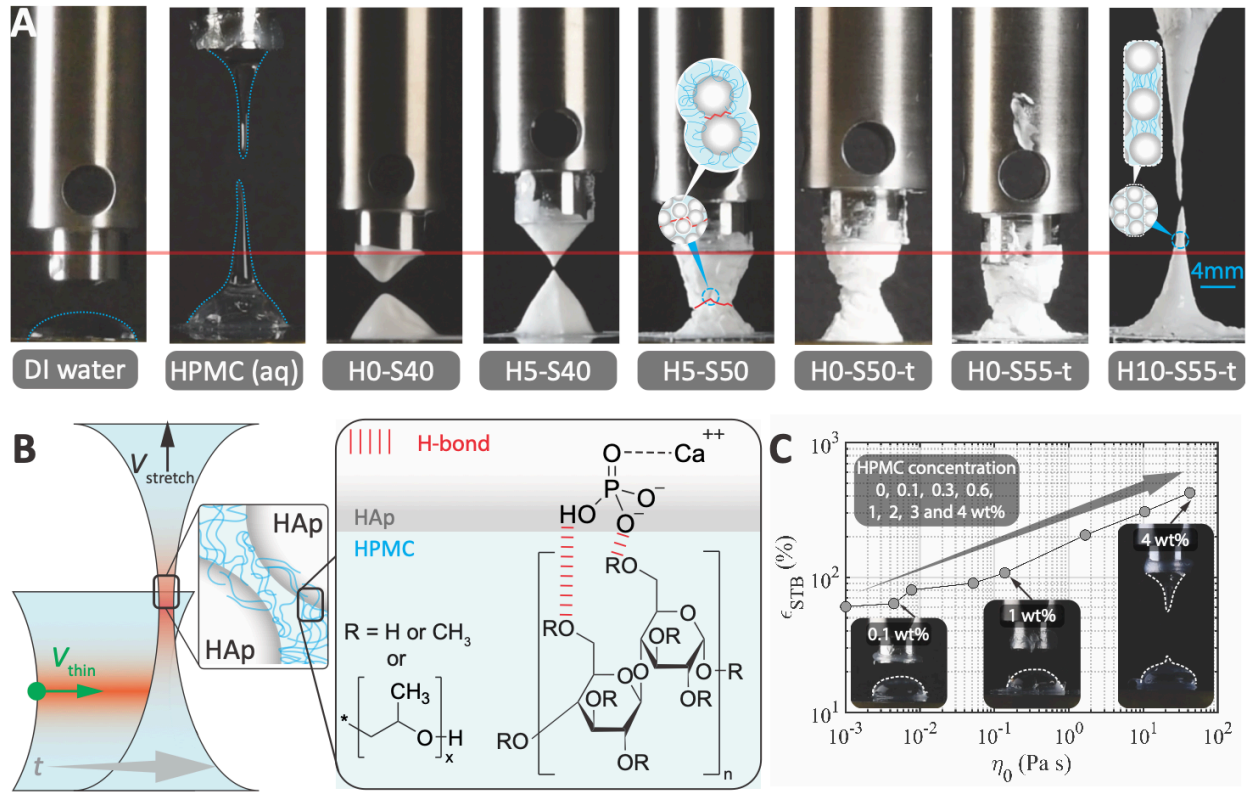

**Figure 3: Material design: microstructure-rheology tuning mechanisms of extensible yield stress fluids. A.** FiSER images at $t = t_b$ = 2.43, 20.16, 5.59, 7.47, 4.60, 6.62, 5.61, 28.8 seconds for water, 5wt% HPMC, and some representative inks respectively. Particle and HPMC additions increase extensibility to a limit. Electrostatic modulation disperses particles surpassing this limit by allowing homogeneous HPMC bridging (depicted in **B**). Red horizontal line shows gap at break of H5-S50 for reference across all samples. **B.** Extensional strain-at-break ($\varepsilon_{STB}$) depends on $V_{stretch}$ (controlled) and $V_{thin}$, influenced by fluid properties (surface tension, $\Gamma$, and density, $\rho$), geometry (initial height, $H$, and diameter, $D$), and rheology (relaxation time, $\tau$, characteristic viscosity, $\eta$, and yield stress $\sigma_y$). HPMC forms polymer-bridged networks with HAp particles *via* hydrogen bonding. **C.** zero-shear viscosity of HPMC solution as a function of concentration, with Oh ~ 0.8 to 800, showing the transition from inertia-capillary to viscous-capillary. The concentration range corresponds to the continuum phase of the HA-based inks in this study.

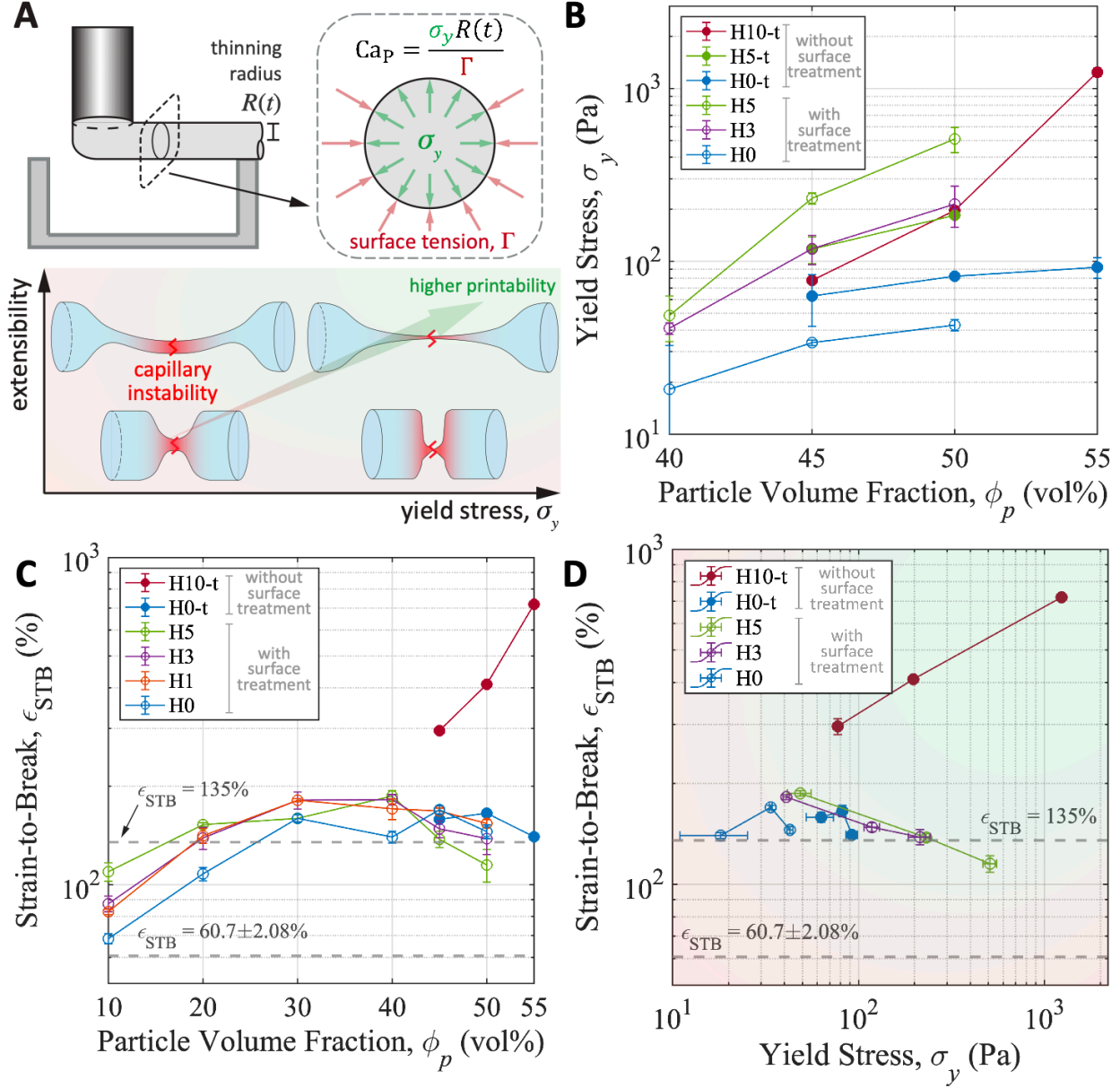

**Figure 4: Microstructure-rheology relationships: the two key rheological properties govern breakup behaviors.** **A.** plasto-capillary number, $\text{Ca}_P \equiv \sigma_y R(t)/\Gamma$, determines smallest filament diameter before pinch-off.[41,57] High extensibility and yield-stress inks excel in DIW. **B.** yield stress, $\sigma_y$, and **C.** strain-to-break, $\epsilon_{STB}$, in material formulation domain. **D.** Ashby plot ($\epsilon_{STB}$ *vs* $\sigma_y$) distinguishes the carefully tuned HAp-based inks from nominal trend. Water capillary stable limit: $60.7 \pm 2.08\%$ (measured) and $135\%$ (theory at zero-gravity[58]). HAp:PMMA is 1:1 vol% for all formulations. Error bars: one standard deviation. Figure S14: Comparison of our inks with other materials in $\sigma_y - \epsilon_{STB}$ space.

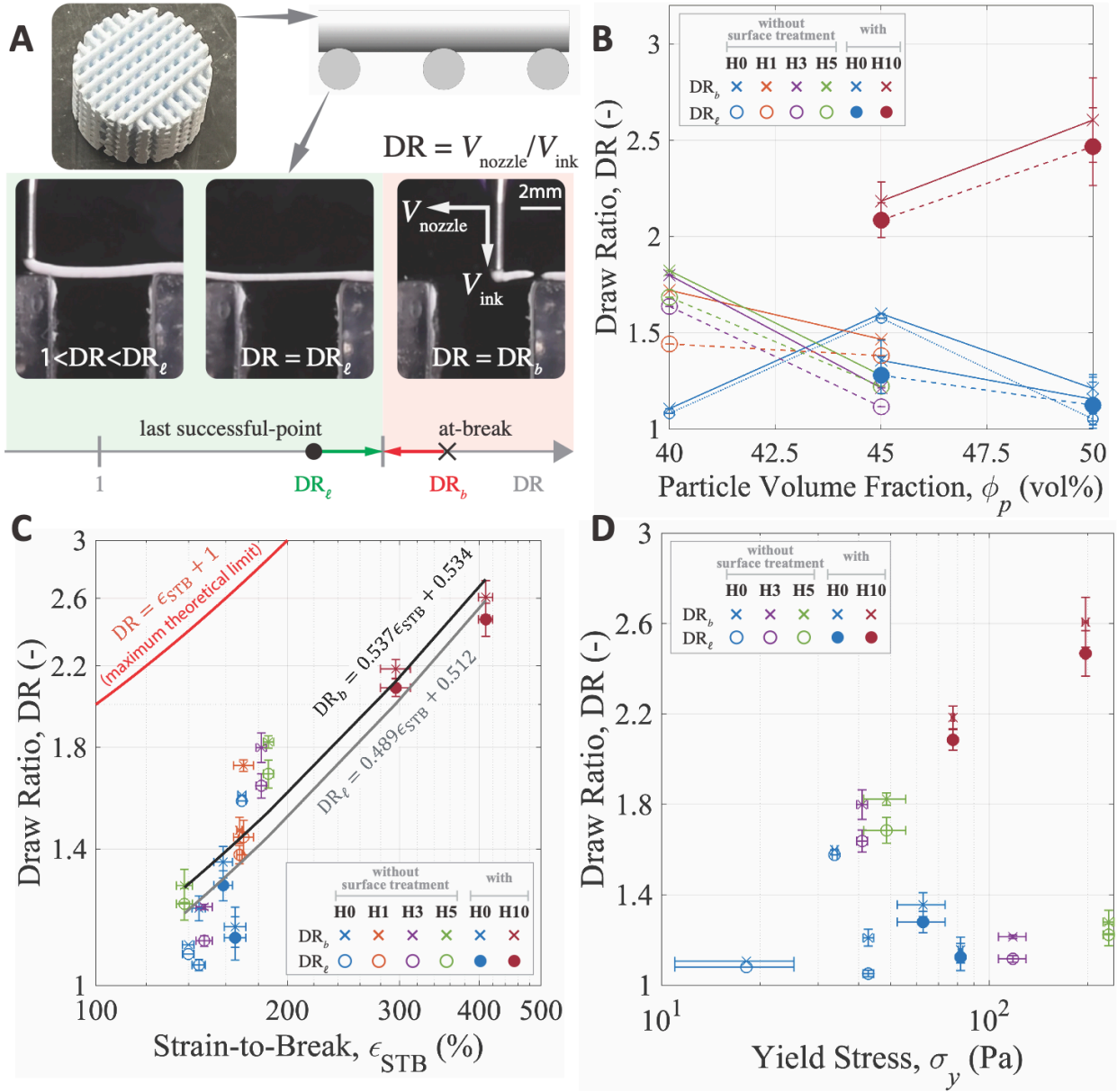

**Figure 5: Rheology-drawability relationships. A.** Gap-spanning experiment showing cases where $1 > \mathrm{DR} > \mathrm{DR}_\ell$, $\mathrm{DR} = \mathrm{DR}_\ell$ and $\mathrm{DR} = \mathrm{DR}_b$. Examples from H10S50-t, representing ~100 tests across formulations with varying $\phi_{\mathrm{HPMC}}$, $\phi_p$, with or without PAA-PEI surface treatment, $V_{\mathrm{ink}}$ and $V_{\mathrm{nozzle}}$. **B.** Draw ratios ($\mathrm{DR}_\ell$, $\mathrm{DR}_b$) in material formulation domain, trend similar to $\epsilon_{\mathrm{STB}}$. **C.** Strong correlation between $\epsilon_{\mathrm{STB}}$ and draw ratios ($\rho_P = 0.93, 0.92$). Red line: $\mathrm{DR} = \epsilon_{\mathrm{STB}} + 100\%$ (Eq. (1)). Black lines: extensibility-drawability relationships (Eq. (2)). **D.** Weak correlation between $\sigma_y$ and draw ratios ($\rho_P = 0.27, 0.26$). HAp:PMMA is 1:1 vol% for all formulations. Error bars: one standard deviation. Supplementary videos show the details of the gap spanning experiments.